\documentstyle[11pt,twoside]{article}

\begin{document}

\newcommand{\dd}{\,{\rm d}}
\newcommand{\ie}{{\it i.e.},\,}
\newcommand{\etal}{{\it et al.\ }}
\newcommand{\eg}{{\it e.g.},\,}
\newcommand{\cf}{{\it cf.\ }}
\newcommand{\vs}{{\it vs.\ }}
\newcommand{\zdot}{\makebox[0pt][l]{.}}
\newcommand{\up}[1]{\ifmmode^{\rm #1}\else$^{\rm #1}$\fi}
\newcommand{\dn}[1]{\ifmmode_{\rm #1}\else$_{\rm #1}$\fi}
\newcommand{\upd}{\up{d}}
\newcommand{\uph}{\up{h}}
\newcommand{\upm}{\up{m}}
\newcommand{\ups}{\up{s}}
\newcommand{\arcd}{\ifmmode^{\circ}\else$^{\circ}$\fi}
\newcommand{\arcm}{\ifmmode{'}\else$'$\fi}
\newcommand{\arcs}{\ifmmode{''}\else$''$\fi}
\newcommand{\MS}{{\rm M}\ifmmode_{\odot}\else$_{\odot}$\fi}
\newcommand{\RS}{{\rm R}\ifmmode_{\odot}\else$_{\odot}$\fi}
\newcommand{\LS}{{\rm L}\ifmmode_{\odot}\else$_{\odot}$\fi}

\newcommand{\Abstract}[2]{{\footnotesize\begin{center}ABSTRACT\end{center}
\vspace{1mm}\par#1\par
\noindent
{\bf Key words:~~}{\it #2}}}

\newcommand{\TabCap}[2]{\begin{center}\parbox[t]{#1}{\begin{center}
  \small {\spaceskip 2pt plus 1pt minus 1pt T a b l e}
  \refstepcounter{table}\thetable \\[2mm]
  \footnotesize #2 \end{center}}\end{center}}

\newcommand{\TableSep}[2]{\begin{table}[p]\vspace{#1}
\TabCap{#2}\end{table}}

\newcommand{\TableFont}{\footnotesize}
\newcommand{\TableFontIt}{\ttit}
\newcommand{\SetTableFont}[1]{\renewcommand{\TableFont}{#1}}

\newcommand{\MakeTable}[4]{\begin{table}[htb]\TabCap{#2}{#3}
  \begin{center} \TableFont \begin{tabular}{#1} #4 
  \end{tabular}\end{center}\end{table}}

\newcommand{\MakeTableSep}[4]{\begin{table}[p]\TabCap{#2}{#3}
  \begin{center} \TableFont \begin{tabular}{#1} #4 
  \end{tabular}\end{center}\end{table}}

\newenvironment{references}%
{
\footnotesize \frenchspacing
\renewcommand{\thesection}{}
\renewcommand{\in}{{\rm in }}
\renewcommand{\AA}{Astron.\ Astrophys.}
\newcommand{\AAS}{Astron.~Astrophys.~Suppl.~Ser.}
\newcommand{\ApJ}{Astrophys.\ J.}
\newcommand{\ApJS}{Astrophys.\ J.~Suppl.~Ser.}
\newcommand{\ApJL}{Astrophys.\ J.~Letters}
\newcommand{\AJ}{Astron.\ J.}
\newcommand{\IBVS}{IBVS}
\newcommand{\PASP}{P.A.S.P.}
\newcommand{\Acta}{Acta Astron.}
\newcommand{\MNRAS}{MNRAS}
\renewcommand{\and}{{\rm and }}
\section{{\rm REFERENCES}}
\sloppy \hyphenpenalty10000
\begin{list}{}{\leftmargin1cm\listparindent-1cm
\itemindent\listparindent\parsep0pt\itemsep0pt}}%
{\end{list}\vspace{2mm}}

\def\TYLDA{~}
\newlength{\DW}
\settowidth{\DW}{0}
\newcommand{\dw}{\hspace{\DW}}

\newcommand{\refitem}[5]{\item[]{#1} #2%
\def\REFARG{#3}\ifx\REFARG\TYLDA\else, {\it#3}\fi
\def\REFARG{#4}\ifx\REFARG\TYLDA\else, {\bf#4}\fi
\def\REFARG{#5}\ifx\REFARG\TYLDA\else, {#5}\fi.}

\newcommand{\Section}[1]{\section{#1}}
\newcommand{\Subsection}[1]{\subsection{#1}}
\newcommand{\Acknow}[1]{\par\vspace{5mm}{\bf Acknowledgments.} #1}
\pagestyle{myheadings}


\def\thefootnote{\fnsymbol{footnote}}


\begin{center}
{\Large\bf The Distance to Pleiades}
\vskip3pt
{\bf B~o~h~d~a~n~~P~a~c~z~y~{\'n}~s~k~i}
\vskip6mm
{Princeton University Observatory, Princeton, NJ 08544-1001, USA\\
e-mail: bp@astro.princeton.edu}
\end{center}
\vskip1cm
\Abstract{
The distance to Pleiades remains controversial.  There is a simple way
to resolve the dispute definitely by measuring the distance to one of
its brightest members, Atlas, which is astrometric and spectroscopic binary.
}

\Section{Introduction}

The distance to Pleiades became a controversial issue ever since Hipparcos
(ESA 1997, Perryman et al. 1997) provided distance measurement to this cluster
(van Leeuwen and Ruiz 1997, Pinsonneault, M. H., et al. 1998), as the new
distance modulus was 5.3 mag rather than the traditional value of 5.6 mag.
This implies that either something is wrong with the commonly used main
sequence fitting method, or systematic errors in Hipparcos parallaxes are
comparable to random errors, i.e. much larger than advertised.  Either
solution is uncomfortable for one or another group of astronomers. 
Percival et al. (2003) present the most recent attempt to modify the main
sequence fitting to agree with Hipparcos parallaxes, and they provide
references to past attempts.  The most serious challenge to Hipparcos 
parallax to Pleiades was made by Narayanan and Gould (1999), who combined 
Hipparcos proper motions with known radial velocities, and using the 
conventional moving cluster method obtained Pleiades distance modulus 5.6 mag. 

Fortunately, there is a simple classical method to resolve the dispute.
Atlas, one of the brightest stars in Pleiades, with $ V = 3.6 $ is known
to be a binary.  Hipparcos:

\begin{figure}[t]  
\vspace{9.0cm}
\includegraphics{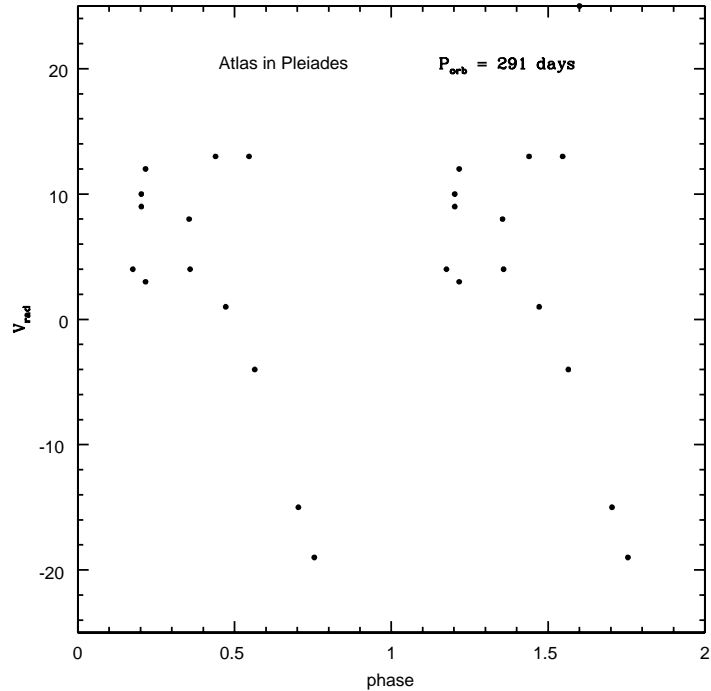}
\caption{
Radial velocities compiled by Abt et al. (1965) for Atlas are plotted
as a function of orbital phase, with orbital period given by Hipparcos.
}
\end{figure}

\centerline{http://astro.estec.esa.nl/Hipparcos/HIPcatalogueSearch.html}

\noindent
gives astrometric orbital period as 290.6598 days, based on a periodic
motion of the light centroid.  The binary has been resolved by the Palomar
Testbed Interferometer and the Mark III Optical Interferometer (Pan et al.
1999).  The astrometric orbit has a semi-major axis of $ 12.9 \pm 0.1 $ mas
and eccentricity of $ 0.246 \pm 0.006 $.  The magnitude difference between
the two components was measured to be 1.7 mag in V band.  In their abstract
Pan et al. (1999) write: `This leads to a distance for Atlas which is
significantly larger than the value of $ 116 \pm 3 $ pc determined by the
Hipparcos for the Pleiades cluster'.  Unfortunately, there was no good 
radial velocity curve available at the time.

The only radial velocity data readily available are those compiled by
Abt et al. (1965), and they are plotted with the Hipparcos period in Fig. 1.
Presumably, these refer to the brighter of the two components.  Fortunately,
modern digital detectors and advanced methods of data analysis (Ruci\'nski
2002) should provide very accurate radial velocity curves for both components
as the star is bright and the observed radial velocity amplitude is large.
Combining astrometric orbit with spectroscopic orbit is the most robust,
purely geometrical method to measure distances with high accuracy and no
serious systematic errors.  This is a classical method.  It was used, for
example, to measure accurate distance to Hyades prior to Hipparcos (e.g. 
Torres et al.  1997).  A recent very impressive application of
this method provided the most accurate distance to the Sgr $ {\rm A^*} $:
$ 8.0 \pm 0.4 $ kpc (Eisenhauer et al. 2003).

It is not clear why the determination of accurate astrometric orbit of Atlas
(Pan et al. 1999) was not followed with a 1\% accurate determination of
the distance to Pleiades.  No matter what the outcome, this would resolve
one of the more dramatic controversies in modern astrophysics, with
important implications either for the main sequence fitting method, or 
for future accurate space astrometry.

\Acknow{
It is a great pleasure to acknowledge discussions with Dr. L. Eyer and
Dr. M. Konacki}



\begin{references}

\refitem{Abt, H. A., Barnes, R. C., Biggs, E. S., and Osmer, P. S.}{1965}{\ApJ}{142}{1604}
\refitem{Eisenhauer, F. et a.}{2003}{astro-ph/0306220}{}{}
\refitem{ESA}{1997}{Hipparcos and Tycho Catalogs, ESA SP-1200 (Nordvijk: ESA)}{Vol. 3}{}
\refitem{Nrayanan, V. K., and Gould, A.}{1999}{\ApJ}{523}{328}
\refitem{Pan, X. P., Shao, M., and Kulkarni, S.}{1999}{AAS Meeting 195}{132.03}{}
\refitem{Percival, S. M., Salaris, M., Kilkenny, D.}{2003}{\AA}{400}{541}
\refitem{Perryman, M. A. C. et al.}{1997}{\AA}{323}{L49}
\refitem{Pinsonneault, M. H., et al.}{1998}{\ApJ}{504}{170}
\refitem{Ruci\'nski, S.}{2002}{\AJ}{124}{1746}
\refitem{Torres, G., Stefanik, R. P., and Latham, D. W.}{1997}{\ApJ}{474}{256}
\refitem{Van Leeuwen, F., and Ruiz, C. S. H.}{1997}{in `Hipparcos', Venice, ed. B. Battrick and M. A. C. Perryman (Noordwijk: ESA)}{689}{}

\end{references}
\end{document}